\documentclass[prb,twocolumn,showpacs,preprintnumbers]{revtex4}

\usepackage{amsmath,amssymb,amsfonts}
\usepackage{dcolumn}
\usepackage{bm}
\usepackage{graphicx}

\begin{document}

\title{Curie temperature  and quantum phase transitions in the Hubbard model
with binary alloy disorder}  
\author{Krzysztof Byczuk$^{1}$ and Martin Ulmke$^{2}$ }

\affiliation{
\centerline {$^1$ Institute of Theoretical Physics,
Warsaw University, ul. Ho\.za 69, PL-00-681 Warszawa, Poland}
\centerline{$^2$
  FGAN - FKIE, Neuenahrer Stra\ss{}e 20, D-53343 Wachtberg, Germany}
 }
\date{\today}

\begin{abstract}
Magnetic and electric properties of the  Hubbard model with binary alloy
disorder are studied within the dynamical mean--field theory.
A paramagnet--ferromagnet phase transition and a Mott--Hubbard
metal--insulator transition are observed upon varying the alloy concentration.
A disorder induced enhancement of the Curie temperature is demonstrated 
and explained by the effects of band splitting and subband filling.  
Different quantum phase transitions driven by changes of the alloy
concentration are identified. 
\end{abstract}

\maketitle

\section{Introduction}

There is a great concern  about the nature
of itinerant ferromagnetism in correlated
electron systems with disorder, such as, e.g., doped manganites
(La$_{1-x}$Sr$_x$MnO$_3$),\cite{imada98,dagotto01} 
alloyed ruthenates (SrRu$_{1-x}$Mn$_x$O$_3$),\cite{cao04}
or alloyed Kondo insulators  (FeSi$_{1-x}$Ge$_x$).\cite{yeo03,anisimov02} 
Moreover,  it
is of fundamental importance for industrial  applications to precisely
control the Curie temperature of different  ferromagnetic  alloys or
diluted  magnetic semiconductors, e.g. ${\rm Ga}_{\rm{1-x}}{\rm
  Mn}_{\rm x}{\rm   As}$.\cite{spintronics,dietl02}

In typical transition metals (e.g. Fe, Ni or Co) the
subtle competition between kinetic and Coulomb energies together with
the  Pauli principle leads to the occurrence of
a ferromagnetic phase.  Since the transition happens when both
contributions to the  total  energy are of the same order, 
theoretical methods to study such systems have to be
nonperturbative.\cite{vollhardt00} 
Within the dynamical mean--field theory
(DMFT),\cite{georges96,pruschke95,vollhardt93} which is genuine   
nonperturbative, detail conditions for occurrence of ferromagnetism
in a one--band Hubbard model were elucidated.\cite{ulmke98,wahle98}
The same nonperturbative scheme together  with a density functional
theory in the local density approximation was used to describe
ferromagnetic phases of Fe and Ni.\cite{lichtenstein01}
The presence of a binary alloy
disorder introduces a new  nonperturbative aspect into the problem.
Namely,  when the difference between ion energies $\Delta$ is larger then the 
band--width $W$, the conducting band is split and two alloy subbands are
formed.\cite{velicky68}

Recent investigation of the one--band Hubbard model
with a  binary
alloy disorder demonstrated an intriguing interplay between effects
due to interaction and randomness.\cite{byczuk03}
It was   shown in  Ref.~[\onlinecite{byczuk03}] that
for electron densities $n <x$, where $x$ was a fixed concentration of an alloy
ion, and  for large local Coulomb 
interactions $U$  the Curie temperature $T_c$  is enhanced when 
$\Delta$ is increased.  Additionally, at  $n=x$ the Mott--Hubbard
metal--insulator transition (MIT) 
occurs for $\Delta = \Delta_c(U)$.   The Mott--Hubbard  transition
in a non--integer filled system with binary alloy disorder  was 
investigated further in 
Ref.~[\onlinecite{byczuk04}] where the notions 
of an \emph{alloy Mott insulator} ($\Delta>U$) and an \emph{alloy
charge--transfer insulator} ($\Delta<U$) were introduced.

In the present paper we study the one--band Hubbard model and determine
its physical properties, such as the Curie temperature, as a function  of
$x$. 
The Mott--Hubbard MIT  can be reached 
by varying $x$ while $n$, $\Delta$ and $U$ are constants. 
Such MIT will be called an \emph{alloy concentration controlled} Mott
transition.  
This notion  extends the previous classification of \emph{band--width
  controlled} and  
\emph{filling controlled} Mott transitions.\cite{imada98}
As will be presented below, under certain circumstances the Curie
temperature has a maximum at a finite alloy concentration.
Additionally, new quantum phase transitions between disorder ferromagnetic and
disorder paramagnetic ground states in the vicinity of the 
 Mott insulators are identified.\cite{belitz} 
These quantum phase transitions are driven by varying the alloy
concentration while the electron density is constant. 
All those effects, as explained below, are caused by 
a subtle interplay between alloy band splitting  and correlations.

The present work is motivated by real experimental situations where
the tuning parameter is  rather $x$ than  $\Delta$.
Namely, while the latter 
is fixed for given atoms the former can be 
varied by making different alloy compositions or
changing  chemical stoichiometry.    Thereby, the controlling of
Curie temperature $T_c(x)$ is experimentally accessible. 
The examples are bcc Fe-Co or fcc Ni-Cu alloys.\cite{bardos69,turek94} In the
first case, due to alloying, the system is driven from weak to strong
ferromagnetism, 
whereas in the latter case the system is changed from a 
ferromagnet to a paramagnet. Interestingly, the behavior of the
saturated magnetization and the Curie temperature in Fe-Co are
non-monotonous  functions of $x$ reaching  maxima at  $30\%$
concentration of Co.\cite{bardos69}  
Another interesting examples are pyrite 
alloys  T$_{1-x}$T'$_x$S$_2$, where
T (T') stands  for transition metals: Fe, Co, Ni, Cu, Zn.\cite{disulfides}
Starting from FeS$_2$ and  alloying it with Co,  the empty $e_g$ band
(FeS$_2$ is a narrow-band semiconductor) is progressively filled in
with electrons and the system becomes  a disordered itinerant
ferromagnet.  Surprisingly, the maximum $T_c(x)$ in
Fe$_{1-x}$Co$_x$S$_2$ occurs  not at $x=1$ but at  $x\approx0.76$.  
In different compound UCu$_2$Si$_{2-x}$Ge$_x$ the
Curie temperature has a maximum at   $x\approx 1.6$, i.e. again when
the system 
is disordered.\cite{neto03}  Interesting aspect of this alloy is that
since Si and Ge are  isovalent, the system is isoelectronic  and the
disorder has only structural character  not influencing  directly the
magnetic arrangement.    

In Section II we introduce a one--band Anderson--Hubbard model with
quenched binary alloy disorder and solve it within DMFT framework. In
Section III we present our results on ferromagnetic properties of this
model when the density  of electrons is independent of 
alloy
concentration, i.e. the system is isoelectronic when $x$ is varied.
Next, in Section IV 
we discuss our results when  the system is non--isoelectronic under the
change of $x$. Section V presents conclusions and a 
final discussion.

\section{Model and Dynamical Mean--Field Theory}

\subsection{Anderson--Hubbard Hamiltonian}

Itinerant electron ferromagnetism in
binary alloy  is described hereby the Anderson-Hubbard
Hamiltonian with uncorrelated on-site disorder
\begin{equation}      
H=\sum_{ij,\sigma }t_{ij}\hat{c}_{i\sigma }^{+} 
\hat{c}_{j\sigma}^{\phantom{+}}+
\sum_{i\sigma} 
\epsilon_i \hat{n}_{i\sigma} + U\sum_{i}\hat{n}_{i\uparrow
}\hat{n}_{i\downarrow },  \label{1} 
\end{equation}
where $t_{ij}$ is the hopping matrix element, $U$ is the local  Coulomb
interaction, $\hat{c}^{+}_{i\sigma}$ 
is the fermionic creation
operator for an electron with spin $\sigma$ in Wannier state $i$, and
$\hat{n}_{i\sigma}$ is the particle number operator. The quenched disorder is
represented by the atomic energies $\epsilon _{i}$, which are random
variables. We consider binary alloy disorder where the atomic energy is
distributed according to the binomial probability density  
\begin{equation}
P(\epsilon_i)=x\delta \left(\epsilon_i
  +\frac{\Delta}{2}\right)+(1-x)\delta \left(\epsilon_i
-\frac{\Delta}{2}\right). 
\end{equation}
Here $\Delta $ is the energy difference between the two atomic 
energies, providing a measure of the disorder strength, while $x$ and
$1-x$ are the concentrations of the two  alloy atoms.  For $\Delta \gg
W$, where $W$ is the  band-width,  it is known that binary alloy
disorder causes a band splitting.\cite{velicky68,commentkb1} The number of
states in each alloy 
subband is equal to $2 xN_a$ and $2(1-x)N_a$, respectively, where
$N_a$ is the number of lattice sites and a factor two counts the  spin
degeneracy. \\

\subsection{Dynamical mean--field theory}

The Anderson--Hubbard Hamiltonian (\ref{1}) is not solvable at any finite 
space dimension. It is however numerically solvable in an infinite 
dimension after a proper rescaling of the hopping parameters,\cite{metzner89}
i.e. a set of self--consistent  DMFT equations is
derived.\cite{georges96,pruschke95,vollhardt93}  The 
local nature of the theory implies that short-range order in position
space is missing. However, 
dynamical correlations due to the local
interaction and disorder\cite{VV+JV92,ulmke95} are fully taken into account. 

In the DMFT scheme the local Green function $G_{\sigma n}$ is given by
the bare density of states  $N^{0}(\epsilon )$ and the local
self-energy  $\Sigma _{\sigma n}$ as 
\begin{equation}
G_{\sigma n}=\int d\epsilon \frac{N^{0}(\epsilon
)}{i\omega _{n}+\mu -\Sigma _{\sigma n}-\epsilon }.
\end{equation}
Here the subscript $n$ refers to the Matsubara frequency
$\omega_{n}=(2n+1)\pi /\beta $ 
for the temperature $T$, with
$\beta=1/k_{B}T$, and $\mu $ is the chemical potential. Within DMFT
the local Green function $G_{\sigma n}$ is determined self-consistently by
\begin{equation} 
G_{\sigma n}=-\Bigg\langle
\frac{\int D\left[ c_{\sigma },c_{\sigma }^{\star }\right]
c_{\sigma n}c_{\sigma n}^{\star }e^{{\cal A}_i\{c_{\sigma
},c_{\sigma }^{\star }, {\cal G}_{\sigma }^{-1}\}}}{\int D\left[
c_{\sigma },c_{\sigma }^{\star } \right] e^{{\cal A}_i\{c_{\sigma
},c_{\sigma }^{\star },{\cal G}_{\sigma }^{-1}\}}} \Bigg\rangle
_{\rm dis}, \label{4}
\end{equation}
together with the \textbf{k}-integrated Dyson equation
\begin{equation}
\mathcal{G}_{\sigma n}^{-1}=G_{\sigma n}^{-1}+\Sigma _{\sigma
n}.
\end{equation}
The single-site action $\mathcal{A}_{i}$ for a site with the ionic
energy $\epsilon _{i}=\pm \Delta /2$ for $i=$A and B, respectively,
has the form 
\begin{widetext}
\begin{eqnarray}
\mathcal{A}_{i}\{c_{\sigma },c_{\sigma }^{\star },\mathcal{G}_{\sigma
}^{-1}\} =\sum_{n,\sigma }c_{\sigma n}^{\star }\mathcal{G}_{\sigma
n}^{-1}c_{\sigma n}-\epsilon _{i}\sum_{\sigma }\int_{0}^{\beta }d\tau
n_{\sigma }(\tau )  
-\frac{U}{2}\sum_{\sigma }\int_{0}^{\beta }d\tau c_{\sigma }^{\ast }(\tau
)c_{\sigma }(\tau )c_{-\sigma }^{\ast }(\tau )c_{-\sigma }(\tau ),  \label{6}
\end{eqnarray}
\end{widetext}
where we used a mixed time/frequency convention for Grassmann
variables  $c_{\sigma }$, $c_{\sigma }^{\star }$.  In the presence of
binary alloy disorder, 
the single impurity problem has to be solved
twice in each self-consistency loop. Averages over the randomness are
obtained by\cite{ulmke95}  
\begin{equation}\langle \cdots \rangle _{\mathrm{dis}}=\int d\epsilon
P(\epsilon )(\cdots ),
\end{equation}
which is equivalent to treating a problem of disorder 
within the coherent potential 
approximation scheme.\cite{velicky68,VV+JV92}
Due to the local nature of the theory and the arithmetic averaging of the 
physical one--particle quantities, Anderson localization 
is not captured.\cite{Anderson58,commentkb2} 

An asymmetric density of states is known to stabilize ferromagnetism in the
one-band Hubbard model for moderate values of 
$U$.\cite{ulmke98,wahle98,vollhardt00}
Therefore, we use the density of states of the fcc-lattice
in infinite dimension,\cite{muller-hartmann91} 
\begin{equation}
N^{0}(\epsilon
)=\frac{\exp [-\frac{1+\sqrt{2}\epsilon }{2}]}{\sqrt{\pi (1+\sqrt{2}\epsilon )}}.
\end{equation}  
This density of states has a square root singularity at  $\epsilon =-1/\sqrt{2}$ and
vanishes exponentially for $\epsilon \rightarrow\infty $.  In the
following, 
the second moment of the density of states is used as the energy
scale and is normalized to unity.
Since the same density of states has been used 
earlier in Refs.~[\onlinecite{ulmke98,byczuk03}]
we are able  to compare the specific numerical 
results.

The one-particle Green function in Eq.~(\ref{4}) is determined by
solving the DMFT equations iteratively \cite{ulmke98,wahle98} using
Quantum Monte-Carlo  simulations  with the Trotter slice $\Delta
\tau=1/4$.\cite{hirsh86} 
Since we are mostly interested in qualitative behavior of $T_c$ vs. $x$ at
different $U$, $n$, and $\Delta$ we do not perform the extrapolation of the
results to $\Delta \tau \to 0$. 
Curie temperatures are obtained by the
divergence of the homogeneous  magnetic
susceptibility explicitly implying that the ferromagnetic phase transitions 
are of the second order.\cite{ulmke98,byczuk01}

\section{Results for isoelectronic alloys}

\subsection{Curie temperature}

\begin{figure}[tbp]{\vspace*{1cm}}
\centerline{\includegraphics [clip,width=7cm,angle=-0]{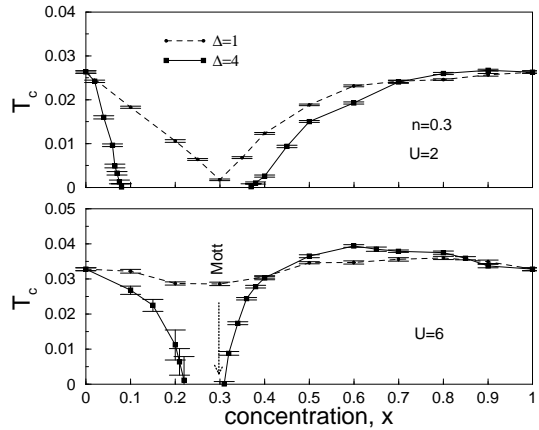}}
\caption{Curie temperature as a function of alloy concentration $x$ at $U=2$ 
(upper panel) and $6$ (lower panel) for $n=0.3$ and disorder $\Delta=1$ (dashed lines) 
and $4$ (solid lines).}
\label{fig1}
\end{figure}

\begin{figure}[tbp]{\vspace*{1cm}}
\centerline{\includegraphics [clip,width=7cm,angle=-0]{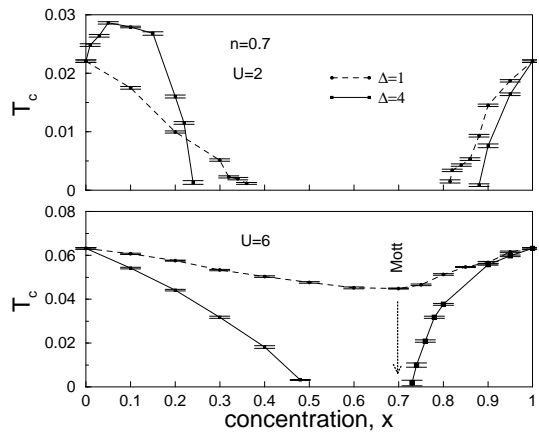}}
\caption{Curie temperature as a function of alloy concentration $x$ at $U=2$ 
(upper panel) and $6$ (lower panel) for $n=0.7$ and disorder $\Delta=1$ (dashed lines) 
and $4$ (solid lines).}
\label{fig2}
\end{figure}

The Curie temperature as a function of alloy concentration
  exhibits very rich and interesting behavior as is documented in
Figs.~\ref{fig1}~and~\ref{fig2}.  It is usually expected that $T_c(x)$ is
suppressed in disordered systems. 
This indeed is found  for most cases when $x$ is varied between 
zero and one.
However, at some concentrations and certain values of 
$U$, $\Delta$ and $n$, the Curie  temperature is enhanced 
above the corresponding value for the 
non-disordered case ($x=0$ or $1$). 
This is shown in both panels of Fig.~\ref{fig1}
for $0.4\lesssim x\lesssim 0.9$ and in the upper panel of
Fig.~\ref{fig2} for $0\lesssim x\lesssim 0.2$. The relative increase
of $T_c$ can be as large as $25\%$,  as is found for $x\approx0.1$ at $n=0.7$,
  $U=2$ and $\Delta=4$ 
(upper panel of Fig.~\ref{fig2}).  

\begin{figure}[tbp]
\centerline{\includegraphics [clip,width=7cm,angle=-0]{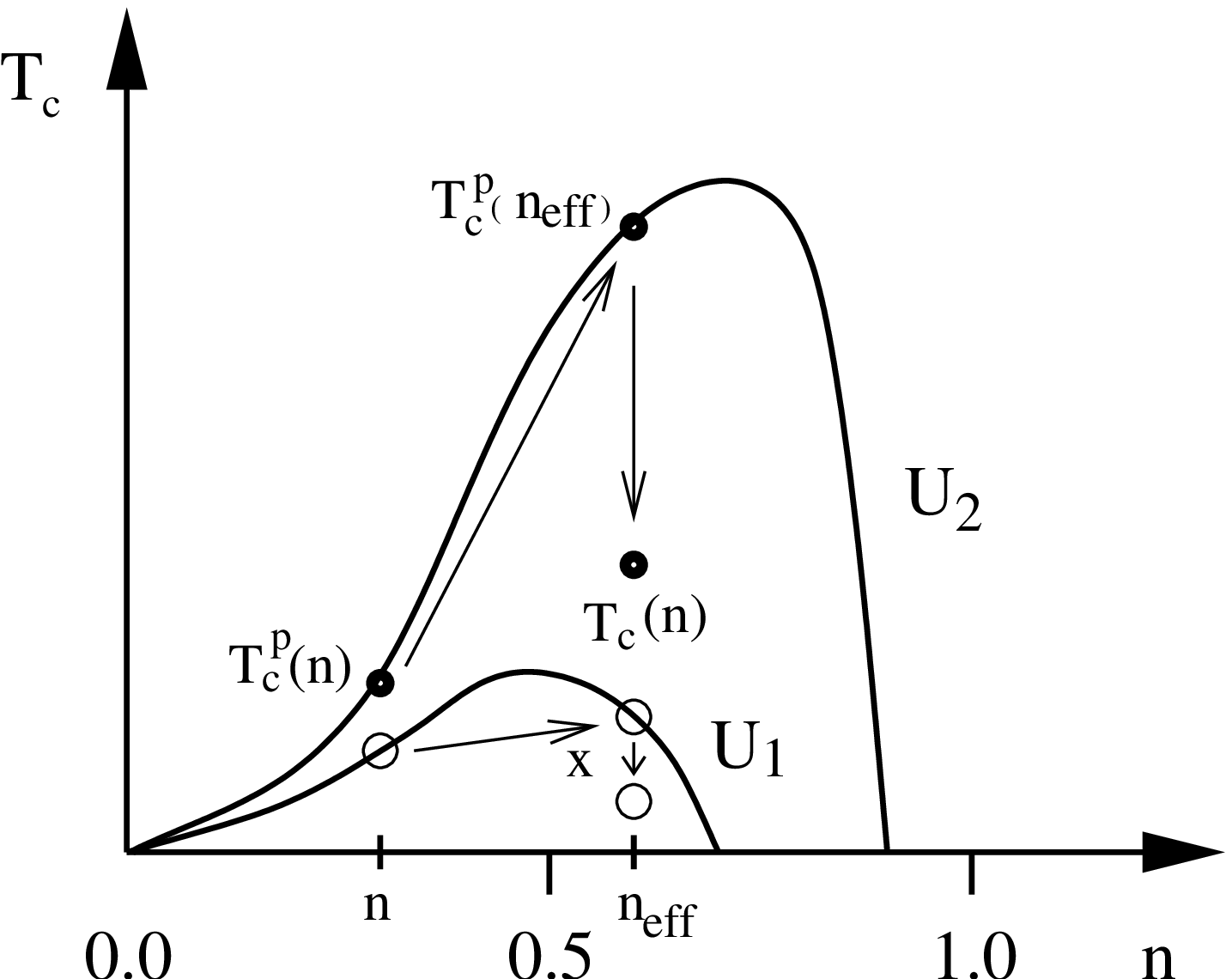}}
\vspace*{1cm}
\centerline{\includegraphics [clip,width=7cm,angle=-0]{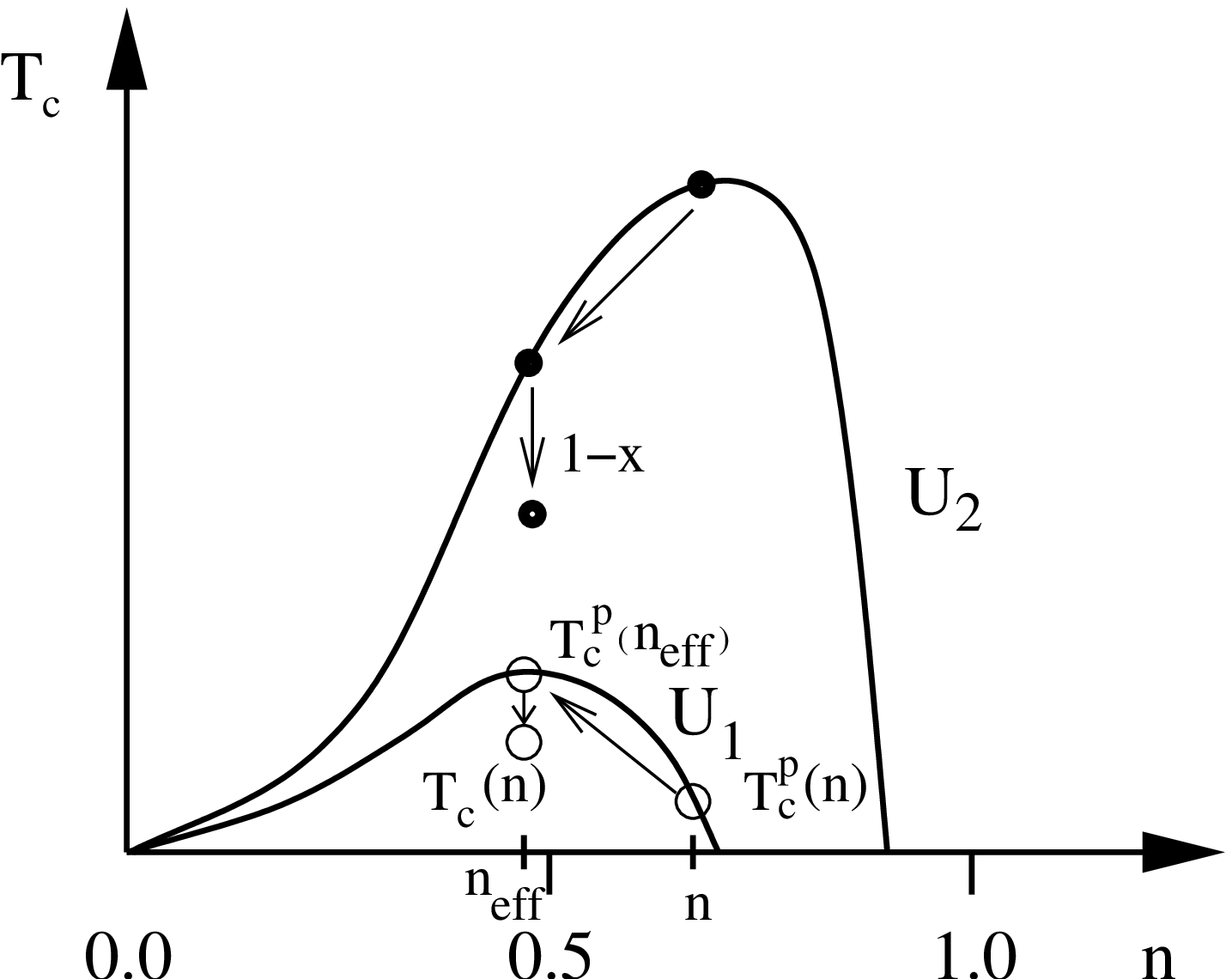}}
\caption{Schematic plots explaining the filling dependence of $T_c$
  for interacting electrons with strong binary alloy disorder. Curves
  represent $T_c^{\mathrm{p}}$, the Curie temperature for the pure
  system, as a function of filling $n$ at two different interactions
  $U_1\ll U_2$.\cite{ulmke98} Upper panel: For $n\lesssim x$, $T_c$ of
  the disordered system can be obtained by transforming the open (for
  $U_1$) and the filled (for $U_2$) point from $n$ to
  $n_{\mathrm{eff}}=n/x$, and then multiplying $T_c^{\mathrm{p}}(n/x)$
  by $x$ as indicated by arrows. One finds
  $T_c(n)<T_c^{\mathrm{p}}(n)$ for $U_1$, but
  $T_c(n)>T_c^{\mathrm{p}}(n)$ for $U_2$.  Lower panel: For $n\gtrsim
  x$, $T_c$ of the disordered system can be obtained by transforming
  $T_c^p(n)$ from $n$ to $n_{\mathrm{eff}}=(n-2x)/(1-x)$, and then
  multiplying $T_c^{\mathrm{p}}[(n-2x)/(1-x)x]$ by $1-x$ as indicated
  by arrows. One finds $T_c(n)>T_c^{\mathrm{p}}(n)$ for $U_1$, but
  $T_c(n)<T_c^{\mathrm{p}}(n)$ for $U_2$.} 
\label{fig4}
\end{figure}

This unusual enhancement of $T_c$ is caused by three distinct features of
interacting electrons in the presence of binary alloy disorder:\cite{byczuk03}

i) The Curie temperature in the non-disordered case 
$T_c^{\mathrm{p}}\equiv T_c(\Delta=0)$, 
depends non-monotonically on band filling $n$.  Namely,
$T_c^{\mathrm{p}}(n)$ has a maximum at some filling $n=n^{\ast}(U)$,
which increases as $U$ is increased;\cite{ulmke98}  see also our
schematic plots in Fig.~\ref{fig4}. 

ii) As was described  above, in the alloy disordered system the band
is split when $\Delta\gg W$.  As a consequence, for $n<2x$  and $T\ll
\Delta$ electrons  occupy only the lower alloy subband and for $n>2x$
both the lower and upper alloy subbands are filled. In the former case
the upper subband is empty while in the later case the lower  subband
is completely full. Effectively, one can therefore describe this system
by a Hubbard model  mapped onto the either lower or the upper alloy
subband, respectively. The second subband plays a passive role.
Hence, the situation corresponds to a \emph{single} band with the
\emph{effective} filling $n_{\mathrm{eff}}=n/x$ for 
$n<2x$ and
$n_{\mathrm{eff}}=(n-2x)/(1-x)$ for $n>2x$.  It is then possible to
determine $T_c$ from the phase diagram of  the Hubbard model without
disorder.\cite{ulmke98}

iii) The disorder leads to a reduction of
$T_c^{\mathrm{p}}(n_{\mathrm{eff}})$ by a factor $\alpha=x$ if the
Fermi level  is in the lower alloy subband or $\alpha=1-x$ if it is in
the upper alloy subband,  i.\,e.\ we find
\begin{equation}
T_c(n)\approx \alpha
T_c^{\mathrm{p}}(n_{\rm eff}),  \label{equation}
\end{equation}
when $\Delta \gg W$ (C.f. Appendix).  Hence, as illustrated in
Fig.~\ref{fig4}, $T_c$ can be determined by
$T_c^{\mathrm{p}}(n_{\mathrm{eff}})$.  Surprisingly, then, it follows
that, for suitable  $U$ and $n$   Curie temperature of a disordered
system can be higher than that of the corresponding non-disordered 
system  [cf.~Fig.~\ref{fig4}].

\begin{figure}[t]{\vspace*{1cm}}
\centerline{\includegraphics [clip,width=7cm,angle=-0]{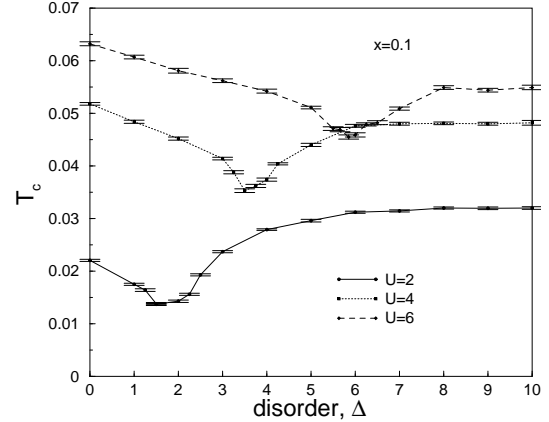}}
\caption{Changes of the Curie temperature with disorder $\Delta$ at
  $x=0.1$ and $n=0.7$ for  different interactions $U$. }
\label{fig3}
\end{figure}

The explanation of the 
$T_c(x)$ enhancement, given above, is supported by a 
detailed analysis how $T_c$ changes when $\Delta$ increases at fixed $x$. 
The numerical results are shown in Fig.~\ref{fig3} at $x=0.1$ and $n=0.7$;
examples 
at $x=0.5$ already have been 
presented in Ref.~[\onlinecite{byczuk03}]. 
For $n>2x$ (results in Fig.~\ref{fig3} corresponds to this regime) 
the Curie temperature initially decreases
upon increasing $\Delta$ from zero. However, when $\Delta \gtrsim U$
 the trend is inverted and $T_c$ increases, finally
saturating. At $\Delta\sim U$ the alloy band splitting becomes effective, 
changing the behavior of $T_c$ versus $\Delta$.
As shown in Fig.~\ref{fig3}, 
only for small $U$ the Curie temperature is 
elevated above the value at $\Delta=0$. This is strongly related to 
the non-monotonic dependence of $T_c^p(n)$, and in particular to the fact 
that its maximum changes with $U$. 
Namely, as is illustrated in the lower panel of 
Fig.~\ref{fig4}, only at small interactions $T_c^p(n_{\rm eff})>T_c^p(n)$,
which is 
a necessary condition for the enhancement of $T_c$ by disorder.
In the $n<2x$ case, on the other hand, the necessary condition 
$T_c^p(n_{\rm  eff})>T_c^p(n)$
for an enhancement of $T_c$ implies that the interaction must be
strong.\cite{byczuk03}

\subsection{Magnetization and Curie constant}

The two different cases $n<2x$ and $n>2x$ (when $\Delta\gg W$) 
should correspond to two different behaviors of a
saturated magnetization, i.e.\ the magnetization for $T\to 0$. 
Namely, in the first case only the lower alloy subband is 
occupied and the magnetization 
$M\equiv \langle \hat{n}_{\uparrow}\rangle-\langle
\hat{n}_{\downarrow}\rangle=n$, where  $\langle \hat{n}_{\uparrow,\downarrow}\rangle$ 
are the average numbers of electrons with spin up or  down.
However, in the second case
$M=n-2x$ since the lower alloy subband is split off and fully occupied by the
electrons with  
two spin species, thereby being magnetically neutral.

\begin{figure}[tpb]{\vspace*{1cm}}
\centerline{\includegraphics [clip,width=7cm,angle=-0]{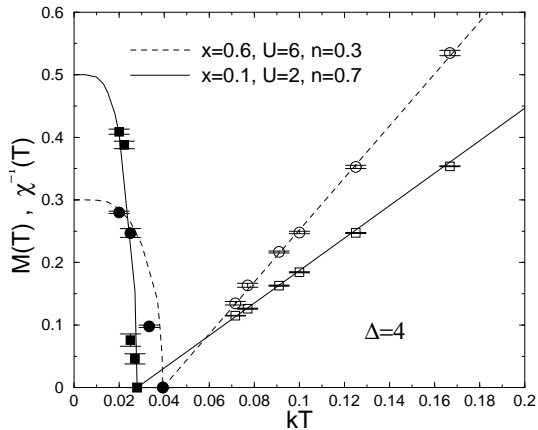}}
\caption{Magnetization and inverse 
static susceptibility at: 
$x=0.6$, $n=0.3$ and $U=6$ (circles and dashed lines), and $x=0.1$, $n=0.7$ and $U=2$ 
 (squares and solid lines). 
In both cases $\Delta=4$.}
\label{fig7}
\end{figure}

In order to confirm this physical picture we calculate the magnetization
$M(T)$ as a function of temperature.   The resultant magnetizations together with the
inverse static susceptibilities $\chi^{-1}(T)$ are  presented in Fig.~\ref{fig7} for
two cases: i) $x=0.6$, $n=0.3$, and $U=6$ (filled and open circles),
corresponding to $n<2x$ case, and ii)
$x=0.1$, $n=0.7$, and $U=2$  (filled and open squares) corresponding to $n>2x$ instance.
At both parameter sets  $T_c$
is enhanced  by  the alloy band splitting ($\Delta=4$).  The numerically 
calculated magnetization  (filled circles and squares)
follow very closely  the theoretical Brillouin curves (dashed and solid
lines).\cite{commentkb3} 
The magnetization data, shown in Fig.~\ref{fig7},
 are consistent with our conjecture that the saturated magnetization should be:  
$M=n=0.3$ in case (i), whereas  $M=n-2x=0.5$ in case (ii).

Two interesting observations are made: 
Firstly, in the case (ii) (with $n>2x$), the presence of disorder increases 
$T_c$ while the saturated
magnetization is suppressed below its value at $x=0$. In other words,
the disordered system becomes a weaker ferromagnet but with higher $T_c$.
Secondly, although the correlated electrons in the disordered system are itinerant, 
the magnetization $M(T)$ is well reproduced by the Brillouin curve, albeit formally 
this curve is derived for localized moments.\cite{ulmke98} The last  observation 
calls for an analytical proof within DMFT analogous to that already 
given for the linear behavior of the inverse susceptibility.\cite{byczuk01} 

Within DMFT the uniform spin susceptibility for a systems with paramagnet--ferromagnet 
phase transition obeys the Curie--Weiss law
$\chi(T)=C/(T-T_c)$,\cite{byczuk01} where $C$ is  
a Curie constant. The same Curie--Weiss law is also derived with a mean--field theory for 
localized magnetic moments and in this case the Curie constant is directly
proportional to the  
saturated magnetization. 
To check whether similar relation holds for itinerant electron system within DMFT 
we compute the Curie constants
obtaining: $C_1=0.240\pm0.002$ in the case (i), and $C_2=0.385\pm0.001$ in the case (ii). 
It turns out that 
the ratio $C_1/C_2=0.623\pm0.005$ is very close to the ratio 
of saturated magnetizations $0.3/0.5=0.6$. 
This result provides a new interpretation of the Curie constant within DMFT, 
i.e. $C$ is proportional to the  saturated 
magnetization, similarly as in localized magnetic moment theory. 
In addition this finding corroborates our initial conjecture regarding
saturated magnetization  
in two physically different cases $n<2x$ and $n>2x$ at large $\Delta$.

\subsection{Metal--insulator transition}

Upon increasing $U$ and $\Delta$, 
the Mott--Hubbard MIT occurs at the 
electronic filling $n=x$.\cite{byczuk03,byczuk04}
Such MIT can also be encountered by varying $x$. This MIT 
will be called   
an \emph{alloy concentration controlled  Mott transition}.
Approaching a  correlated insulator  the itinerant
ferromagnetism is also suppressed due to localization of the electrons. 
Using linearized DMFT\cite{bulla00} it was estimated\cite{byczuk04} that when
$\Delta\to \infty$  
the critical interaction $U_c\approx 6\sqrt{x}$.
Such estimation leads to  $U_c\approx 3.3$ for $n=0.3$ (Fig.~\ref{fig1}), and
to  $U_c\approx 5$ for $n=0.7$ (Fig.~\ref{fig2}). 
This means that at $\Delta=4$ and $U=6$ (lower panels in
Figs.~\ref{fig1}~and~\ref{fig2})  
the Mott--Hubbard MIT is possible at $x=n$.

\subsection{Quantum phase transitions}

At large $U$ and $\Delta$ the system  encounters several quantum ($T=0$) phase
transitions when $x$ is varied. 
One is the Mott--Hubbard MIT discussed above. 
The other is the transition from a ferromagnetic metal to a paramagnetic metal. 
In this quantum phase transition the Curie temperature vanishes at
$x=x_{c}$ which is different from $x=n$, where Mott--Hubbard MIT occurs. 

The presence  of the quantum phase transitions 
 should affect  the properties of the disordered Fermi
liquid at finite temperatures, provided that the system is of finite
dimensionality. 
This scenario would be worth to be investigated including spatial correlations,
which however goes beyond the present DMFT framework. 

\section{Non-isoelectronic alloy}

\begin{figure}[pbt]{\vspace*{1cm}}
\centerline{\includegraphics [clip,width=7cm,angle=-0]{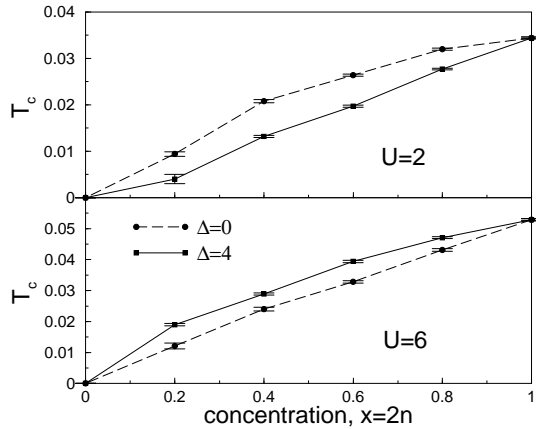}}
\caption{Behavior of the Curie temperature for $U=2$ (upper panel) and
$6$ (lower panel) when the change of electron filling $n$ is associated
with the change in the alloy concentration $x$. At large $U$ 
the Curie temperature is enhanced in the disorder system.}
\label{fig5}
\end{figure}

In many alloys the change in the concentration $x$ alters 
the electron density  $n$.  In this Section we  investigate $T_c(x)$ 
when $x$ and $n$ are varied simultaneously. The results presented in
Fig.~\ref{fig5} are obtained under the 
assumption that  $x=2n$. When $x=0$ the
band is empty (the system is a band insulator). Upon increasing $x$ from zero
to one, 
the band filling  increases from zero to one--half (quarter filled band). 
This model realization  would correspond to the physical situation in
Fe$_{1-x}$Co$_x$S$_2$ 
alloy if the number of states in $e_g$ band is normalized to one. 
Of course, this analogy should not be stressed 
too far since in our model 
an important exchange interaction (Hund's coupling) is absent. 

As shown in Fig.~\ref{fig5}  the presence of disorder at weak interaction
always suppresses the Curie temperature with respect to the 
non--disorder
case.  However, at large $U$ and $\Delta\gtrsim 1$ the Curie temperature
is larger than that in the pure case at the same filling. 

This difference again can be understood on the basis provided by the scheme
depicted in Fig.~\ref{fig4}.
In the present case  $n<2x$ for all $x$, and only the 
lower alloy subband 
plays a role in the effective description at large $\Delta$.  
In this limit this subband is  effectively  filled with  $n_{\rm eff}=n/x=1/2$ electrons. 
Using the same arguments as in the upper panel in Fig.~\ref{fig4} we see that $T_c(x)$ is 
enhanced only for large $U$. \\

\section{Conclusions}

In the present paper we studied the one--band Anderson--Hubbard model with
binary alloy disorder showing that the Curie temperature in such alloyed
correlated electron system can 
reach higher values than those in the non-disorder system.
We also identified and
discussed the metal--insulator  transition at non-integer
fillings. Additionally possible quantum phase transitions at zero temperature
were described pointing out on a possible interesting quantum critical
behavior of a disordered Fermi liquid. 
This work completes previous studies of itinerant 
ferromagnetism in the pure and disordered Hubbard model within
DMFT.\cite{ulmke98,wahle98,byczuk03}

Regarding the physical systems, it is now of great importance to
extend the one--band Hubbard model with alloy disorder to a multi--band case
and formulate the adequate version of DMFT for solving this
problem. It is very interesting which aspect of the ferromagnetism in
one--band alloy are generic and will be present in multi--band
case. \\

\begin{acknowledgments}
We thank R.~Bulla and D.~Vollhardt 
for useful discussions and Ch.~Leighton for communication on
Fe$_{1-x}$Co$_x$S$_2$ pyrite. 
This work was supported in part by the Sonderforschungsbereich 484 of
the Deutsche Forschungsgemeinschaft. We thank the Institute of Physics
of the University of Augsburg for hospitality and for being able to make
use of their computer resources.
Financial support of KB through KBN-2 P03B 08 224 is also acknowledged. 
\end{acknowledgments}

\appendix*
\section{Hartree--Fock theory}
Within the Hartree-Fock approximation one can find analytically that in the strong 
disordered limit $\Delta \gg W$ the self--energy has the form
\begin{equation}
\Sigma_{n\sigma}=\sigma \frac{U M}{4} +(1-2x)\frac{\Delta}{2}+ 
\frac{x(1-x)\Delta^2}{z-\frac{\sigma U M}{4}-\frac{2-x}{2}\Delta},
\end{equation}
where $M$ is the magnetization density. This self--energy leads to the splitting 
of the density of states with $x$ and $1-x$ of the initial states in each alloy subband.
Since the integrated function in the equation for $T_c^{\rm HF}$ is 
peaked 
at 
$\omega=\mu$, only one of the subband gives contribution to evaluate $T_c^{\rm HF}$.
As a result, $T_c^{\rm HF}\approx \alpha T_c^{{\rm HF}\;p}$, where $\alpha=x$ or $1-x$, as 
introduced in Section II. 
Although in DMFT one cannot find analytically the corresponding self--energy, the splitting 
of the density of states also appears and only one of the subbands contributes in the 
equation for $T_c$. 
In analogy to the Hartree--Fock approximation, we  assume that $T_c$ is reduced by $\alpha$ with 
respect to $T_c^p$, which we find to be valid even at strong
interaction.


\begin{references}
\bibitem{commentkb1} The alloy band splitting occurs in any dimension. However, a 
genuine disorder induced gap in the spectrum appears only if the bare density of 
states has a finite support. Otherwise a soft-gap is expected.\cite{byczuk03}
\bibitem{commentkb3} The Brillouin curves follow from  solutions of the self--consistent equation
$M(T)/M_s=\tanh[T_cM(T)/(TM_s)]$, where $M_s$ are assumed to be saturated magnetizations and
$T_c$ are earlier computed Curie temperatures.
\end{references}
\end{document}